\documentclass[journal]{IEEEtran}

\pdfoutput=1 

\usepackage[noadjust]{cite}
\usepackage{pslatex} 
\usepackage{booktabs}

\usepackage{graphicx}
\usepackage{xcolor}
\usepackage{multirow}
\usepackage{amsmath}
\usepackage{amssymb}
\usepackage{amsfonts} %
\usepackage{calligra} %
\usepackage{mathtools}

\usepackage{amsthm} 

\usepackage{dblfloatfix} 

\usepackage{tikz}
\usepackage{verbatim}
\usetikzlibrary{shapes,backgrounds}
\usetikzlibrary{arrows,fit,automata,positioning,decorations,calc}
\usetikzlibrary{spy}
\usetikzlibrary{matrix,chains,decorations.pathreplacing}

\usepackage{caption}
\usepackage{subcaption} 

\usepackage{comment}
\usepackage{listings}

\usepackage{pgfplots}
\usepackage{pgfplotstable}

\usetikzlibrary{patterns}

\usepackage[percent]{overpic}

\usepackage[english]{babel}
\usepackage{blindtext}

\usepackage{acronym}

\usepackage{listings}
\usepackage{float}
\usepackage{textcomp}

\theoremstyle{definition}
\theoremstyle{example}
\theoremstyle{theorem}

\usepackage{acronym}
\usepackage{algorithm}
\usepackage{algorithmic}

\usepackage{url} 
\definecolor{mygreen}{rgb}{0,0.6,0}
\definecolor{mygray}{rgb}{0.5,0.5,0.5}
\definecolor{mymauve}{rgb}{0.58,0,0.82}

\lstdefinestyle{VStyle}{ 
  backgroundcolor=\color{white},   
  basicstyle=\scriptsize,        
  breakatwhitespace=false,         
  breaklines=true,                 
  captionpos=b,                    
  commentstyle=\color{mygreen},    
  deletekeywords={...},            
  escapeinside={\%*}{*)},          
  extendedchars=true,              
  frame=single,	                   
  keepspaces=true,                 
  keywordstyle=\color{blue},       
  language=Verilog,                 
  morekeywords={*,...},            
  numbers=left,                    
  numbersep=5pt,                   
  numberstyle=\tiny\color{mygray}, 
  rulecolor=\color{black},         
  showspaces=false,                
  showstringspaces=false,          
  showtabs=false,                  
  stepnumber=1,                    
  stringstyle=\color{mymauve},     
  tabsize=2,	                   
  title=\lstname                   
}

\definecolor{mGreen}{rgb}{0,0.6,0}
\definecolor{mGray}{rgb}{0.5,0.5,0.5}
\definecolor{mPurple}{rgb}{0.58,0,0.82}
\definecolor{backgroundColour}{rgb}{0.95,0.95,0.92}

\lstdefinestyle{CStyle}{
    backgroundcolor=\color{backgroundColour},   
    commentstyle=\color{mGreen},
    keywordstyle=\color{magenta},
    numberstyle=\tiny\color{mGray},
    stringstyle=\color{mPurple},
    basicstyle=\scriptsize,
    breakatwhitespace=false,         
    breaklines=true,                 
    captionpos=b,                    
    keepspaces=true,                 
    numbers=left,                    
    numbersep=5pt,                  
    showspaces=false,                
    showstringspaces=false,
    showtabs=false,                  
    tabsize=2,
    language=C
}

\input{lstlang.sty}
\lstdefinestyle{AVRStyle}{
    backgroundcolor=\color{backgroundColour},   
    commentstyle=\color{mGreen},
    keywordstyle=\color{magenta},
    numberstyle=\tiny\color{mGray},
    stringstyle=\color{mPurple},
    basicstyle=\scriptsize,
    breakatwhitespace=false,         
    breaklines=true,                 
    captionpos=b,                    
    keepspaces=true,                 
    numbers=left,                    
    numbersep=5pt,                  
    showspaces=false,                
    showstringspaces=false,
    showtabs=false,                  
    tabsize=2,
    language=AVR
}
\acrodef{WCRT}{Worst Case Reaction Time}
\acrodef{WCET}{Worst Case Execution Time}
\acrodef{SOT}{Start Of Tick}
\acrodef{EOT}{End Of Tick}

\acrodef{ESS}{Energy Storage System}

\acrodef{CPS}{Cyber-Physical System}
\acrodef{IoT}{Internet of Things}

\acrodef{AM}{Additive Manufacturing}
\newcommand{\ignore}[1]{{}}

\newcommand{\squishlist}{
	\begin{list}{$\bullet$}
		{ \setlength{\itemsep}{0pt}
			\setlength{\parsep}{1pt}
			\setlength{\topsep}{1pt}
			\setlength{\partopsep}{0pt}
			\setlength{\leftmargin}{0.9em}
			\setlength{\labelwidth}{1.5em}
			\setlength{\labelsep}{0.4em} } }
	\newcommand{\squishend}{
	\end{list}  } 

\newcommand{\sol}{FLAW3D}

\begin{document}

\title{FLAW3D: A Trojan-based Cyber Attack on the Physical Outcomes of Additive Manufacturing}
%
\author{Hammond Pearce \emph{(Member, IEEE)},
		Kaushik Yanamandra,
		Nikhil Gupta \emph{(Senior Member, IEEE)},\\
		Ramesh Karri \emph{(Fellow, IEEE)\vspace{-5mm}}
\thanks{H. Pearce is with the Dept. of Electrical and Computer Engineering, New York University, NY 11201 USA (e-mail: hammond.pearce@nyu.edu)}
\thanks{K. Yanamandra is with the Dept. of Mechanical and Aerospace Engineering, New York University, NY 11201 USA (e-mail: vsy212@nyu.edu)}
\thanks{N. Gupta is with the Dept. of Mechanical and Aerospace Engineering, New York University, NY 11201 USA (e-mail: ngupta@nyu.edu}
\thanks{R. Karri is with the Dept. of Electrical and Computer Engineering, New York University, NY 11201 USA (e-mail: rkarri@nyu.edu)}%
}%
\maketitle %
%
%
\begin{abstract}
%
%
%
Additive Manufacturing (AM) systems such as 3D printers use inexpensive microcontrollers that rarely feature cybersecurity defenses.
This is a risk, especially given the rising threat landscape within the larger digital manufacturing domain.
In this work we demonstrate this risk by presenting the design and study of a malicious Trojan (the FLAW3D bootloader) for AVR-based Marlin-compatible 3D printers ($>$100 commercial models). We show that the Trojan can hide from programming tools, and
even within tight design constraints (less than 1.7 kilobytes in size), it can compromise the quality of additively manufactured prints and reduce tensile strengths by up to 50\%.

\end{abstract}


\IEEEpeerreviewmaketitle

\section{Introduction}
\label{sec:intro}

\ac{AM}, also known as \textit{3D printing}, is a technique whereby materials are deposited and fused to produce volumetric parts.
In recent years there have been considerable advances in the field, 
and \ac{AM} is increasingly being adopted across a range of industries (e.g. within aerospace~\cite{nickels_am_2015}, biomedical ~\cite{placone_recent_2018}, 
and others). The advantages are numerous: \ac{AM} allows for the creation of complex and bespoke products without complex tooling, allows for pull-based manufacturing of products on demand rather than in advance, and rapid prototyping to  iterate over product designs.

With the increasing attention on \ac{AM} cyber-physical systems (CPSs), 
there has been an increased scrutiny on the cybersecurity of the production process.
Potential vulnerabilities have been highlighted at every step of the digital and physical supply chains~\cite{gupta_additive_2020,moore_vulnerability_2016,wu_cybersecurity_2018,mahesh_survey_2020}.
The impacts and implications of successful \ac{AM} cyber attacks have been explored, with demonstrations showing that parts can be modified at print using malicious firmware~\cite{moore_implications_2017}. Even subtle modifications can have insidious consequences~\cite{zeltmann_manufacturing_2016} (e.g. defects being introduced in drone propellers causing them to fail prematurely in flight~\cite{belikovetsky_dr0wned_2017}).

However, while these works highlight the potential for exploitation, they do not examine the actual pathways for doing so within an \ac{AM} CPS --
yet in order to craft suitable defenses for attacks on \ac{AM} CPS, we must have an idea of how they might be performed.
In this paper we examine how a malicious modification can be introduced in a 3D printer firmware so as to compromise print quality. 
This is a realistic threat due to the hidden complexity of firmware in printers from the `hobbyist' to the `commercial grade', and potentially malicious and/or insecure code has already been highlighted as a likely attack vector~\cite{moore_implications_2017,wu_cybersecurity_2018}.
While initial work studied common printer firmware (such as in Marlin and Repetier~\cite{moore_vulnerability_2016}) for vulnerabilities, these analyses overlook the elemental piece of the firmware/software stack, the \emph{bootloader}. 

Within the AM context, bootloaders are small  firmware components fundamental to the operation of the printer software. Typically, they are not replaced/updated during the product lifecycle. Bootloaders do two tasks: (1) install the higher-level firmware into the controller memory when requested. (2) launch the installed firmware after a normal power up sequence.
Crucially, bootloaders are generic, and often used across different products -- 
%
a common occurrence within the culture of 3D printer implementations, which often share hardware and software designs (e.g. those popularized via the open-source RepRap project~\cite{jones_reprap_2011}). 
This means a single bootloader Trojan may be utilized to target a large number of AM machines, making it an attractive attack mechanism.

In this work we consider the bootloaders installed into the low-level controllers within commercial 3D printers (where they might be one part of the control system) and in hobbyist 3D printers (where they can be the only controller).
However, though this paper frames and demonstrates an attack around low-end desktop 3D printers, these mechanisms could be used to target any CPS with embedded firmware running on insecure hardware (e.g. PCB printers, IC fabrication and test). 


\subsection{Contributions}

This is the first comprehensive study of a bootloader-based attack on AM CPS. Contributions are 4-fold, and organized in this paper as follows:
\S\ref{sec:background} presents a study of the related work and attack surface for 3D printers given their underlying implementations.
\S\ref{sec:trojan-design} presents a design space exploration of a proof-of-concept firmware Trojan \sol\ (pronounced `flawed') which targets Marlin-compatible AVR controllers in 3D printers.
\S\ref{sec:defects} performs a qualitative and quantitative evaluation of the Trojan by examining two different mechanisms that can compromise print quality, and in \S\ref{sec:prevention} we provide a discussion and walkthrough of how the Trojan could be detected and prevented.
Finally, \S\ref{sec:conclusions} concludes.

\section{Background and Related Work}
\label{sec:background}
%


\begin{table*}[t]
\resizebox{2\columnwidth}{!}{%
\begin{tabular}{@{}lll@{}}
\toprule
\textbf{Reference}           & \textbf{Type}                     & \textbf{Summary}                                                                                                                                                \\ \midrule
Moore et al.~\cite{moore_vulnerability_2016}
& Vulnerability analysis   & Automatic checking of 3D printer firmware for vulnerabilities.                                                                                         \\

Moore et al.~\cite{moore_implications_2017}
& Specific compromise      & Demonstrates the potential for malicious 3D printer firmware to introduce defects into 3D printed parts.                                               \\

Belikovetsky et al.~\cite{belikovetsky_dr0wned_2017}
& Specific compromise      & Automatically adding defects to 3D printed drone blades.                                                                                               \\
ESET~\cite{noauthor_acadmedre_2012}
& Specific compromise      & ACAD/Medre.A worm in AutoLISP steals engineering CAD files.                                                                                            \\
Slaughter et al.~\cite{slaughter_how_2017}
& Specific compromise      & Temperature manipulation reliably introduces defects in metal AM.                                                                                      \\
Zeltmann et al.~\cite{zeltmann_manufacturing_2016}   
& Specific compromise                   & Discusses how subtle defects can be deliberately introduced to 3D prints that reduce print quality.                                                                           \\

OSHA~\cite{noauthor_after_nodate}
& Safety violation example & Company fined for 3D printer workplace safety violations leading to explosion.                                                                         \\
Mahesh et al.~\cite{mahesh_survey_2020}
& Survey                   & Survey of cybersecurity state of the art for AM. Attack / defense modelling, taxonomies, case studies. \\

Prinsloo et al.~\cite{prinsloo_review_2019}
& Survey                   & Security risks within the Industry 4.0 manufacturing domain.                                                                                           \\
Graves et al.~\cite{graves_characteristic_2019}
& Survey                   & Risk survey and attack / defense modelling for AM, noting differences with traditional manufacturing.                                                  \\
Yampolskiy et al.~\cite{yampolskiy_using_2016}   
& Survey                   & Survey and analysis of potential avenues for 3d printers to be 'weaponized'.                                                                           \\

Yampolskiy et al.~\cite{yampolskiy_language_2015}
& Taxonomy                 & Specifying manipulations as tuples of \{influenced elements, influences\} giving \{affected elements, impacts\}.                             \\
Gupta et al.~\cite{gupta_additive_2020}
& Survey / Taxonomy        & Examining the supply chain and build process for AM cybersecurity vulnerabilities.                                                                     \\ \bottomrule
\end{tabular}%
}
\caption{Survey of literature discussing attacks on AM printers.}
\label{tbl:survey}
\vspace{-6mm}
\end{table*}

\subsection{Attacker Motivation and Attack Taxonomies}
The two major motivations for an attack on an \ac{AM} CPS are~\cite{gupta_additive_2020}: (1) IP theft via product reverse engineering and/or counterfeiting, and (2) sabotage of either the printed part or the 3D printer producing them.
Both result in financial outcomes, either in the attacker gaining proprietary (valuable) knowledge, or in reducing the reputation or value of the attacked system. 
Overall in the literature a number of potential and realized attack strategies have been published, with a summary of these presented in Table~\ref{tbl:survey}.
Many of the works focus on the motivations, taxonomy, and theoretical basis for attacks, rather than the specific technical steps required to achieve them.
In addition, where specific attack methodologies are detailed, attack implementation is either considered out of scope or implemented within the higher-level cyber realm (such as within \cite{belikovetsky_dr0wned_2017} and \cite{noauthor_acadmedre_2012}).
Importantly, when considering attacks in this higher-level area, many defenses already exist via standard information and cybersecurity best practices (e.g. ensuring trusted software updates, firewalls, virus scanning, etc.).
The state of the art is less detailed when considering attacks on the \textit{low-level hardware} of 3D printers. 
The two major works in this space come from \cite{moore_vulnerability_2016}, which detailed an exploration of 3D printer firmware vulnerabilities (e.g. to denial of service and data corruption attacks), and in \cite{moore_implications_2017}, which augments 3D printer firmware directly to add malicious code. 
However, while effective, their implementation strategy of their attacks on the 3D printer firmware has several technical shortcomings -- specifically, it relies on simply changing firmware codes directly, re-compiling, and re-downloading.
If a program's source code can be changed in your attack model (e.g. by a malafide insider), then any behavior change is possible.
As a result, it is a primary focus of designers to audit their firmware (as in \cite{moore_vulnerability_2016}) and detect these changes before deployment.

In this paper we wish to expand the vision for low-level attack strategies and motivate the need for more defensive mechanisms within 3D printers. For this, we present an attack strategy which does not rely on changing the 3D printer firmware directly. 
We present this through an examination of a complete attack life-cycle, performing a deep dive into the technical details of how common 3D printers function, and include installation mechanisms, exploitation strategies and triggers, and mechanisms to avoid detection and removal.


\subsection{Software and Hardware Trojans}
In the cybersecurity domain, `Trojan Horses' refer to deliberate fault-like modifications made to a design for malicious aims, either to steal information or to cause system failure~\cite{vosatka_introduction_2018} (aligning them closely with the aforementioned CPS attack motivations!).
Trojans, which may be created by malicious actors working in a product's design team or from compromised CAD tools used during product creation, have three essential characteristics: malicious intent, evasion detection, and activation rarity~\cite{bhunia_hardware_2014}. 
A Trojan may seek to leak cryptographic information / design files, cause digital / real-world damage, reduce operational reliability / product life-time. 
While typically software-based~\cite{landwehr_taxonomy_1994}, there has been recent attention on hardware Trojans embedded in systems, for instance within integrated circuits~\cite{bhunia_hardware_2014} or encoded into PCBs~\cite{ghosh_how_2015}.
Although out of scope for this work, this does raise interesting recursive possibilities within AM for PCBs (i.e. PCB printers~\cite{botfactory_inc_botfactory_nodate}). 
Here, a firmware Trojan in the printer could insert hardware Trojans into produced PCBs.

\subsection{The 3D Printer Attack Surface}
\label{sec:attack-surface}
\begin{table}[h]
\vspace{-3mm}
\resizebox{\columnwidth}{!}{%
\begin{tabular}{@{}l|lllll@{}}
\toprule
\textbf{Printer}            & \textbf{\begin{tabular}[c]{@{}l@{}}Commercial\\ grade?\end{tabular}} & \textbf{``Open''?} & \textbf{\begin{tabular}[c]{@{}l@{}}Marlin\\ compatible?\end{tabular}} & \textbf{\begin{tabular}[c]{@{}l@{}}Native\\networking?\end{tabular}} & \textbf{\begin{tabular}[c]{@{}l@{}}Support F/W\\ updates?\end{tabular}} \\ \midrule
Stratasys Elite             & Y                                                                    & N                & N                                                                  & Y                                                                       & Y                                                                    \\
Ultimaker S5 Pro            & Y                                                                    & N                & N                                                                  & Y                                                                       & Y                                                                    \\
Ultimaker 2+                & Y                                                                    & Y                & Y                                                                  & N                                                                       & Y                                                                    \\
Makerbot Replicator Z18     & Y                                                                    & N                & N                                                                  & Y                                                                       & Y                                                                    \\
Makerbot Replicator (orig.) & N                                                                    & Y                & Y                                                                  & N                                                                       & Y                                                                    \\
Anet A8                     & N                                                                    & Y                & Y                                                                  & N                                                                       & Y                                                                    \\
Printrbot Simple            & N                                                                    & Y                & Y                                                                  & N                                                                       & Y                                                                    \\ \bottomrule
\end{tabular}%
}
\caption{Features of select commercial / hobbyist 3D printers} 
\label{tbl:printers}
\vspace{-2mm}
\end{table}

Thanks largely to the quality and success of the open source 3D printer projects, many 3D printer implementations (hardware and software) are closely related. 
The RepRap 3D printer project~\cite{jones_reprap_2011}, itself based on early industrial Fused Deposition Modelling (FDM) printers, has inspired at least 81 models of 3D printers directly~\cite{noauthor_reprap_nodate} (with the derivatives often going on to inspire further designs).
Table~\ref{tbl:printers} summarizes some examples. 3D printers are a mix of ``open'' (i.e. open schematics, hardware layouts, and software) and ``closed'' (more likely the commercial printers with trade secrets). The (listed) printers all have some method for updating the installed firmware. In general, ``open'' printers are often compatible with third-party firmware -- the most popular being the Arduino-based Marlin~\cite{lahteine_home_nodate}, which supports over 100 printer models~\cite{noauthor_release_nodate}.
While formerly a premium feature, increasing numbers of 3D printers are also beginning to include native support for networking. Where this isn't included, 3D printers may be networked by wiring connections to external print servers, e.g. Octoprint~\cite{hausge_octoprintorg_nodate}. A ``networked'' 3D printer may just include an Octoprint server internally!
Overall, the networking of 3D printers is becoming a concern, especially when they are exposed to the wider internet (e.g. IoT scanning website Censys estimates $>2500$ Octoprint servers exposed globally\footnote{Search using \url{https://censys.io/ipv4?q=octoprint}}).

\begin{figure}[h]
    \centering
    \includegraphics[width=0.45\textwidth]{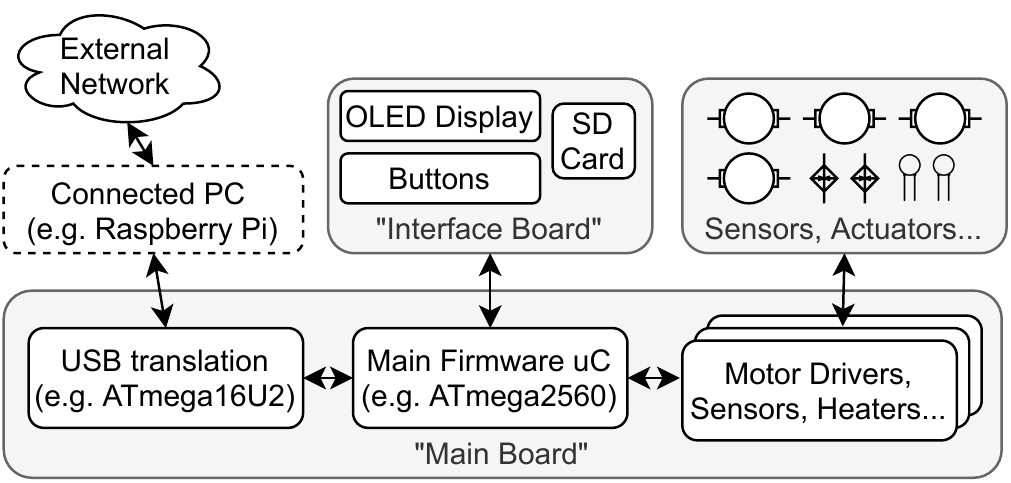}
    \caption{Generic DM architecture for a 3D printer}
    \label{fig:um2-hardware}
    \vspace{-2mm}
\end{figure}

All 3D printers have their low-level hardware control (e.g. control of the sensors and actuators) managed by time-predictable microcontrollers. Computationally intensive functionality such as GUIs and networking (if present) are managed by more powerful embedded or general purpose devices (e.g. as in the Ultimaker S5 Pro) or by external connected devices. 
An example of this kind of hierarchy can be seen in a generalized 3D printer architecture, detailed in Figure~\ref{fig:um2-hardware}.
Here, the high-level functionality (networking, slicing, and so on) is provided by an external general purpose computer - for example, a Raspberry Pi, which could be running software such as Octoprint or Ultimaker Cura.
This is interfaced with the printer control firmware, which is distributed across a network of microcontrollers.
Crucially, unlike in the Raspberry Pi (and other general purpose computers), where mechanisms such as ``Verified Boot''~\cite{rigas_enabling_2019} or ``Secure Boot''~\cite{wilkins_uefi_2013}) can be used to ensure the security of the low-level firmware, there is no specific functionality to perform this in low-end microcontrollers.
%
Security instead becomes the responsibility of the installed bootloader (if deemed necessary) -- which may include features for cryptographically checking the firmware updates before installation~\cite{lau_secure_2012}.
Of course, if the bootloader itself is replaced (either via external hardware circuitry or via the network-connected high-end embedded systems), nothing may verify that the replacement bootloader is free of malice. 
Further, as the bootloader is not part of the firmware (e.g. Marlin), \textit{even if the firmware is audited for security risks, the bootloader may be excluded from this analysis}.
%
%
%
%
%
This is especially a concern in desktop / hobbyist AM, where bootloaders are often provided in binary form (preventing adequate auditing of their source codes) and re-used across many compatible devices. 

\begin{figure}[h]
    \vspace{-3mm}
    \centering
    \includegraphics[scale=0.8]{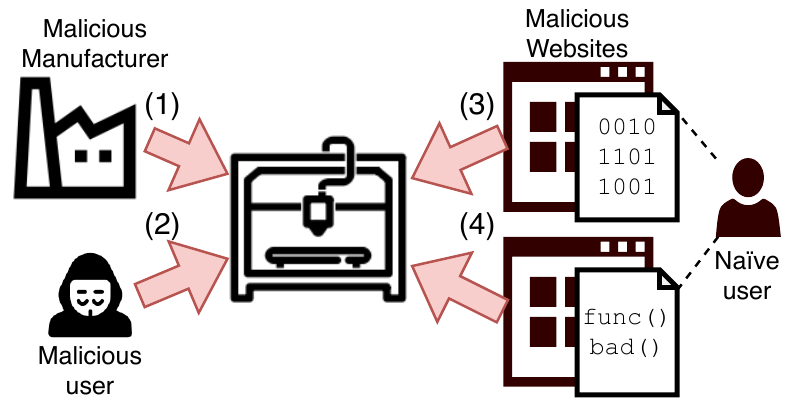}
    \caption{Attack surface for  \sol.}
    \label{fig:attack-surfaces}
    \vspace{-2mm}
\end{figure}

If the adversary has (or has had) physical access to the printer-under-attack or is otherwise able to trick someone with physical access into installing the malicious bootloader, then four attack vectors are possible, as depicted in Fig.~\ref{fig:attack-surfaces}: (1) the original manufacturer of the printer or a malicious insider,  (2) a malicious user with access to the printer, (3) a third-party (e.g. a website) provides a pre-compiled malicious bootloader or (4) a third-party provides malicious bootloader source code and a na\"ive user installs it without adequate auditing.

This risk is expanded by the presence of the \textit{network} of microcontrollers. As can be seen in Fig.~\ref{fig:um2-hardware}, which represents a common architecture used within 3D printers, there are two different microcontrollers which could each individually interfere with the correct operation of the device. 
This only becomes more challenging given commercial and industrial-grade additive manufacturing machines, which can feature tens to hundreds of microcontrollers all running different code.

\section{The \sol\ Bootloader}
\label{sec:trojan-design}
\subsection{Goals}
In this section we discuss the design space exploration for the creation of a new firmware Trojan called \sol. It will target AVR-based desktop 3D printers which run the Marlin printer firmware.  
We note that the attack will not target any specific feature of any printer, rather, it aims to misuse features from the underlying AVR microcontrollers and the operating Marlin firmware. We also note that the general methodology in this section can be used to target other printers, including at the `commercial grade' by malafide insiders.

Overall we will consider an adversary that aims to sabotage a design firm by reducing the quality of printed designs. 
In order to achieve this we seek to give the Trojan the ability to both \textit{relocate} and \textit{remove} printed material.

We must also work within constraints: specifically, the Trojan must not increase the compiled size of the bootloader beyond the boot flash size limit.
For example, in the ATmega2560, a microcontroller commonly used in 3D printers, this is just 8192 bytes, of which 5786 bytes are already taken by the existing bootloader code, leaving approx. 2406 bytes for the Trojan). 

\subsection{Arduino-compatible bootloaders for AVR}

Marlin is installed on AVR-based 3D printers via compatible software (e.g. the Arduino IDE, the Ultimaker Cura slicer) running on a secondary machine (e.g. a Raspberrry Pi, or a general purpose Windows or Linux computer). 
The gatekeeper for this process is the AVR-based bootloader which resides on the target microcontroller.
Upon power-up it executes a simple state machine, which initializes a UART communication peripheral and awaits valid bootloader commands. If no commands arrive before a timeout, the existing main firmware is initialized if available. If a command does arrive, the bootloader will execute it. These commands, which are based on a subset of the STK500 standard~\cite{noauthor_stk500_2006}, include instructions to read and write flash and EEPROM memory. Three observations are important.

\noindent
\textbf{Observation One:} 
though both bootloaders and applications are installed into the microcontroller flash memory, they do not run simultaneously. 
Bootloaders run first, eventually loading the main firmware.
When this happens, bootloaders are entirely unloaded, with the stack and global memory reset and reconfigured for the main application. 

\noindent
\textbf{Observation Two:} for bootloaders to install the main firmware, all memory values for the binary must pass through them both during upload (installation) and download (verification). A secure bootloader could perform cryptographic and data integrity checks, but regular Arduino-compatible bootloaders do not.
Instead, data verification is managed by the off-chip toolchain reading all memory addresses after installation and ensuring they match the expected.

\noindent
\textbf{Observation Three:} though Arduino-compatible bootloaders do not tend to utilize interrupts, the underlying hardware supports this.
Interrupts function by preempting control flow to specific locations in memory (known as interrupt vector tables).
As the bootloader and the main firmware are distinct applications, they must have different vector tables. 
Hence, the AVR architecture supports changing the address of the vector table using a special control register \texttt{IVSEL}.

\subsection{Design of a generic Arduino-compatible Trojan for AVR}
\label{sec:trojan-generic}
Using the bootloader source code provided by Arduino at~\cite{noauthor_arduinoarduino-stk500v2-bootloader_2021} as a start point, we now craft a Trojan for an ATmega2560, noting that the steps for other common AVR microcontrollers are largely the same. 
Firstly, in order for the Trojan to function it needs to be able to inject instructions into the program executed by the main firmware.
Based on \textbf{Observation Three}, this is achieved via the interrupt vector table select register \texttt{IVSEL}.
By default this register is configured to make interrupts jump to the vector table associated with the main firmware. 
If changed, the machine will instead jump to \textit{bootloader} program space upon an interrupt occurring.
Crucially, the startup code (i.e the code that runs `before \texttt{main()}') generated by the AVR compiler \texttt{avr-gcc} \textit{does not check or set the \texttt{IVSEL} register} --- nor does the firmware we are interested in hijacking (Marlin).
This means that if we define our own bootloader interrupt service routines (ISRs), and set \texttt{IVSEL} 
before booting the main application, 
\textit{the bootloader ISRs will replace the main application's}.

However, if the main application defines ISRs, and those ISRs are never called (because the hardware is calling the wrong interrupt vectors) then the presence of the Trojan will be easily noticed.
Thus, the Trojan must embed within its ISRs calls to the main application ISRs (using the addresses of the original vector table).
In this way it \textit{wraps} the application ISRs -- allowing injection of both \textit{prologue} and \textit{epilogue} instructions to each routine.
Code to perform these injections, using the \texttt{avr-gcc} compiler, is presented in Listings~\ref{lst:inject-isr} and \ref{lst:move-ivsel}. 

\begin{figure}[h]
\vspace{-2mm}
\begin{lstlisting}[style=CStyle,caption=Structuring an AVR bootloader ISR to `inject' code,label=lst:inject-isr]
//inject code in this vector
ISR(..._vect) { 

  ... // prologue injection here 

  //call application's vector address then run CLI
  // this is important as the app. ISR will return 
  // using RETI, which re-enables interrupts after
  // the next instruction - a CLI prevents this
  asm volatile("call [..._vect_addr]\n\t" 
               "cli\n\t");  
	
  ... // epilogue injection here
	
} // our ISR will return with another RETI
\end{lstlisting}
\vspace{-5mm}
\end{figure}

\begin{figure}[h]
\vspace{-2mm}
\begin{lstlisting}[style=CStyle,caption=Two-step process changes \texttt{IVSEL} in \texttt{main()},label=lst:move-ivsel]
void main(void) {
  //create a copy of MCUCR
  char temp;
  temp = MCUCR;
  // Enable change of Interrupt Vector location
  MCUCR = temp | (1<<IVCE);
  // Point interrupts at Bootloader Flash section 
  MCUCR = temp | (1<<IVSEL);
    
  ... //rest of the bootloader
}
\end{lstlisting}
\vspace{-8mm}
\end{figure}

While this structure allows for the injection of instructions, declaring state (variables) that will persist outside of the ISRs is a separate, more difficult issue, as the processor memory is reinitialized by the main application (\textbf{Observation One}).
In other words, the Trojan has no safe way of storing global or static variables outside of ISR invocations.

To resolve this we consider the implementation of the AVR's Harvard-style memory architecture.
The data memory, which is separated from the instruction memory, is partitioned into register space (in the first 256 bytes) and general RAM.
In the general RAM, static objects and global variables are placed in the low addresses by the C compiler, and the stack (which stores local variables and function return addresses) grows from the highest address \textit{downwards}.

To keep track of the stack's position, two registers are provided -- \texttt{SPH} and \texttt{SPL} (for the high and low byte of the 16-bit address respectively). 
During the startup code of an AVR application 
one of the first tasks that is performed is re-initialization of these two registers.
An example which sets \texttt{SPH}/\texttt{SPL} to \texttt{0x21}/\texttt{0xFF} can be seen in the ATmega2560 startup disassembly in Listing~\ref{lst:startup-disassembly}.
Note that, given the stack grows downwards, if the values loaded into \texttt{SPH} and \texttt{SPL} are decreased, then the addresses above their new value are excluded from the stack. This would free them for the Trojan.

\begin{figure}[h]
\vspace{-3mm}
\begin{lstlisting}[style=AVRStyle,caption=Disassembled Marlin ATmega2560 startup code,label=lst:startup-disassembly]
00000000 <__vectors>:
  0: 0c 94 e7 10  jmp 0x39dc ; jump <__dtors_end>
  ... ; ... etc      
  
000039dc <__dtors_end>:
  ; clear SREG
  39dc:	11 24  eor  r1, r1
  39de:	1f be  out  0x3f, r1	
  ; code to set SPH/SPL
  39e0:	cf ef  ldi r28, 0xFF ; load 0xFF to R28
  39e2:	d1 e2  ldi r29, 0x21 ; load 0x21 to R29
  39e4:	de bf  out 0x3e, r29 ; set SPH to R29 (0x21)
  39e6:	cd bf  out 0x3d, r28 ; set SPL to R28 (0xFF)
  ... ; ... etc      
\end{lstlisting}
\vspace{-5mm}
\end{figure}

Now, recall {\bf Observation Two:} all instructions making up the main application are passed through the bootloader during installation.
That is, the bootloader is responsible for receiving the compiled application binary over UART and saving it into flash memory. In addition, though the startup code which initializes \texttt{SPH} and \texttt{SPL} may be located unpredictably within the binary, the \textit{specific} instructions that make it up \textit{do not change from application to application}, and further, are usually located at the first jump from the vector at program address \texttt{0000} (the reset vector). 
This means that during the installation loop, the bootloader can scan for the pattern of 8 bytes which sets \texttt{SPH}/\texttt{SPL} (Lines 10-13 in Listing~\ref{lst:startup-disassembly}) and then \textit{alter those bytes that represent the data address before saving the program into application flash}. To minimise detection, the Trojan should change the value only slightly (otherwise the application has a higher chance of running out of memory unexpectedly during operation), so it alters the instruction \texttt{ldi r28, 0xFF} to \texttt{ldi r28, 0xF0} (subtracting 15). This way 15 bytes of data memory are excluded from the stack for use by the Trojan for saving variables to global state.

While the edit to the program binary can be detected in the normal case, now recall again {\bf Observation Two}: specifically that the verification of the binary is also done via reading out the saved binary using the same bootloader. 
This means that by simply performing the edit step in reverse, the bootloader can change the binary values back to the expected when the programming tool attempts to ensure that the program has been uploaded correctly. 
This means that based on the described Trojan design, \textit{programming tools which rely on the bootloader to upload, download, and verify the installed program cannot detect the malicious modifications}.



\subsection{Design of \sol}
\begin{figure}[b!]
\vspace{-3mm}
    \centering
    \includegraphics[width=0.48\textwidth]{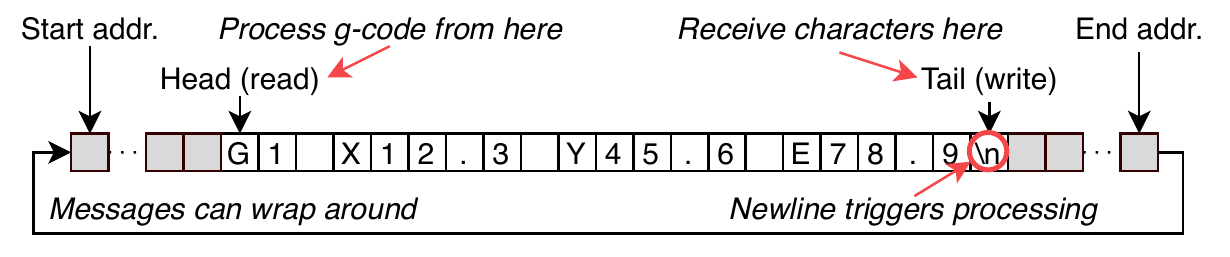}
    \caption{Ring Buffer implementation in Marlin.}
    \label{fig:ring-buffer}
\end{figure}
Returning to the original goal of subvert printing quality, consider now the flow of information in a 3D printer. 
Print commands, which detail control sequences for the motors, extruders, and heaters of the printer, are specified in textual \texttt{g-code} language. These originate from a computer running a slicer program that converts 3D computer aided design models into the \texttt{g-code}. From the point of view of the controller running the printer, the commands arrive character-by-character via the UART peripheral.

It is thus the goal of \sol\ to edit this incoming \texttt{g-code}, using
the Trojan framework from \S\ref{sec:trojan-generic} as the starting point. 
Using the ISR code injection mechanism the reception of valid \texttt{g-code} can be compromised and intercepted `in-flight'. 
The Trojan can then edit received commands before they are processed by the main application. 

While it might appear that this can be done by interfering with the UART peripheral, (e.g. editing the received character during the injection), two hardware constraints prevent this: (1) the UART RX register is \textit{read-only}, and (2) reading from the UART RX register has side effects; once read, it unlocks the hardware for further reception.
We thus instead consider how the bootloader can interfere with the higher level firmware (Marlin).
Marlin, prior to processing the received \texttt{g-code} with the main process loop, uses its UART RX ISR to store received characters in a \textit{ring buffer} (depicted in Fig.~\ref{fig:ring-buffer}). 

As the AVR has no memory protection, the Trojan can access the entire memory space from the compromised ISRs. If the location of the ring buffer can be deduced, the Trojan can read and edit the \texttt{g-code} commands prior to their processing by the main application.
To accomplish this, consider {\bf Observation Two}. As the Trojan can access the flash memory of the microcontroller, it can scan the UART ISR of Marlin, revealing two distinctive \texttt{lds} commands near the start (Listing~\ref{lst:marlin-disassembly}).

\begin{figure}[!t]
\vspace{-2mm}
\begin{lstlisting}[style=AVRStyle,caption={Disassembly of UART RX ISR in Marlin on AVR},label=lst:marlin-disassembly]
00000000 <__vectors>:
  ... ; ... etc    
  50: 0c 94 b7 83  jmp 0x1076e ; jump to UART RX ISR
  ... ; ... etc    
  
0001076e <__vector_20>:
  ... ; ... etc    
  10786: 20 91 24 03  lds r18, 0x0324 
  1078a: e0 91 23 03  lds r30, 0x0323  
  ... ; ... etc      
\end{lstlisting}
\vspace{-8mm}
\end{figure}   

The addresses in these two instructions correspond to the location in memory of the \textit{head} and \textit{tail} pointers of the ring buffer. By default the compiled Marlin firmware's data structure layout will place these pointers 128 bytes after of the ring buffer itself. This means that the smaller of the two addresses (e.g. 0x0323) minus 128 gives the root address in memory of the ring buffer where the incoming \texttt{g-code} is stored. \sol\ thus encodes this behavior as a function \texttt{find\_ring\_buffer()} to do this task automatically prior to launching the main application firmware. 
The function performs this by traversing the binary of the main application, starting from the (constant) UART RX ISR vector location, and following the program jumps and the linear path of execution until it finds these back-to-back \texttt{lds} commands. If it does not identify them within 256 instructions, it aborts, and the Trojan is rendered dormant. 
If it succeeds, it stores the head pointer, tail pointer, and root address in the global state variables that we established earlier (in the top 15 bytes of the AVR memory).
\sol\ can then use these pointers with string manipulation code injected as an epilogue of the main application UART RX ISR: and now, incoming \texttt{g-code} can be edited. 

As standard string manipulation in C (performed by functions such as \texttt{sscanf} to read out variables and \texttt{sprintf} to rewrite them) are too large to use within the context of a bootloader, \sol\ relies on a simple embedded state machine. 
This examines incoming \texttt{g-code} strings character by character, and can internally convert received ASCII-encoded floating point values into integer-type fixed-point notation. 
If the bootloader detects that a target value to edit is arriving, it suppresses the Marlin firmware's normal behavior by editing the ring buffer head pointer addresses to hide the incoming characters.
Then, once the target value has been entirely received, the bootloader can process and edit it before restoring the correct pointer value and allowing Marlin to detect and process the command.

\section{Inducing Defects with \sol}
\label{sec:defects}

\subsection{Overview}
\begin{figure}[b]
    \vspace{-3mm}
    \centering
    \includegraphics[width=0.48\textwidth]{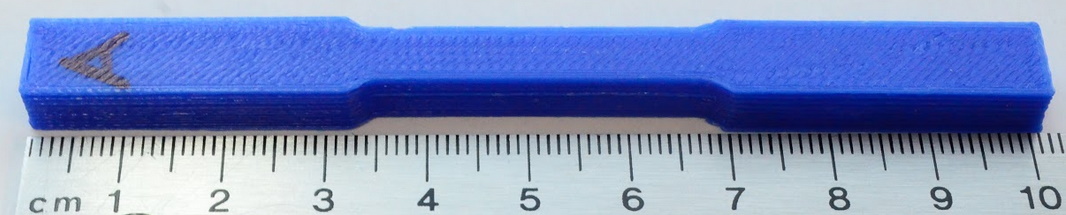}
    \caption{Example control group test specimen.}
    \label{fig:sample-A}
\end{figure}

\sol\ scans and alters incoming \texttt{g-code} before processing by Marlin. Given the restricted space for Trojan code (e.g. $\sim$2406 bytes on ATmega2560), the edits need to be simple. Complex edits may also cause noticeable delays. Given these constraints, we present two Trojan methodologies, with code compiled using \texttt{avr-gcc} version 5.4.0 with optimization \texttt{-Os}. To measure the impact of the Trojans on print quality, strength tests were performed using the tensile test specimen design E8 from ASTM A370-20~\cite{noauthor_astm_2020} (Fig.~\ref{fig:sample-A}).
To check consistency, samples were printed with two different sets of slicing parameters using two different anonymized commercially available AVR-based 3D printers which use Marlin internally  (Table~\ref{tbl:print-details}).
\begin{table}[h]
\resizebox{\columnwidth}{!}{%
\begin{tabular}{@{}ll|llll@{}}
\toprule
\multicolumn{2}{c|}{\textbf{Printer info.}}                              & \multicolumn{4}{c}{\textbf{(Print) Slicing settings}}                                                                                                                                                                                                                        \\ \midrule
ID & \begin{tabular}[c]{@{}l@{}}Cost (New)\\ (USD)\end{tabular} & \begin{tabular}[c]{@{}l@{}}Layer height\\ (mm)\end{tabular} & \begin{tabular}[c]{@{}l@{}}Line width\\ (mm)\end{tabular} & \begin{tabular}[c]{@{}l@{}}Infill\\ strategy\end{tabular} & \begin{tabular}[c]{@{}l@{}}Infill line\\ distance (mm)\end{tabular} \\ \midrule
A  & 2499                                                       & 0.15                                                        & 0.35                                                      & Cubic (18\%)                                              & 5.83                                                                \\
B  & 599                                                        & 0.15                                                        & 0.4                                                       & Grid (10\%)                                               & 8.0                                                                 \\ \bottomrule
\end{tabular}%
}
\caption{Printers and print settings.}
\label{tbl:print-details}
\vspace{-3mm}
\end{table}

While it is customary to report strength in the form of maximum load sustained divided by the cross-section area, as the Trojan modifies the printer \texttt{g-code} the exact cross-sectional area may vary from test to test. Hence, in this paper the normalized maximum load sustained by the specimens before failure is compared, as all specimens have the same origin test geometry. This is performed destructively using an Instron 4467 universal test system. Though it is deterministic, to minimize printer ``noise'' each sample is printed five times and test results averaged. 

\subsection{Trojan Attack 1: Material Reduction}

\begin{figure}[b]
    \centering
    \includegraphics[width=0.48\textwidth]{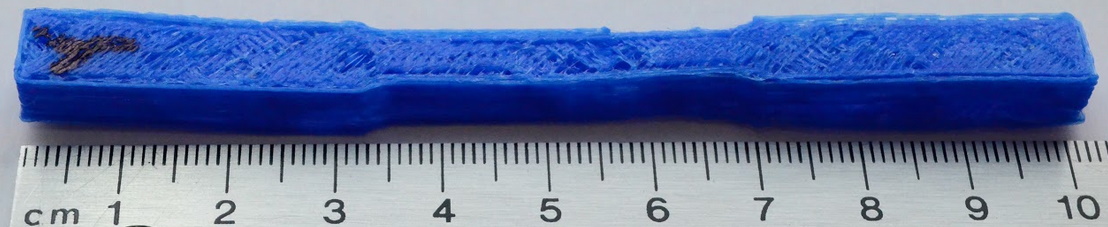}
    \caption{Example 50\% material reduction.}
    \label{fig:sample-Y}
\end{figure}

This Trojan attack reduces the amount of printed material.
\sol\ uniformly scans for the  \texttt{G1} commands (linear move) in the incoming \texttt{g-code} which include the extrusion command (character \texttt{E}). 
Then, the extrusion value is decreased by some percentage. For instance, a command \texttt{G1 X2 Y3 E4}, which moves to (X,Y) (2,3) and extrudes 4mm of filament can be edited to \texttt{G1 X2 Y3 E2} to reduce the material by 50\%. 
While this reduces the maximum tensile strength of the specimen, the attack is easily detectable, both by weight tests and by visual inspection in severe cases (Fig.~\ref{fig:sample-Y}).

The normalized test results are given in Fig.~\ref{fig:pa-material-reduction} and~\ref{fig:pb-material-reduction}. 
As can be seen, material reduction reduces mass and reduces maximum tensile load fairly linearly. 
The attack increases the bootloader size from 5786 bytes to 7422 bytes, an increase of \textbf{1636 bytes}. 

\newlength{\tempdima}

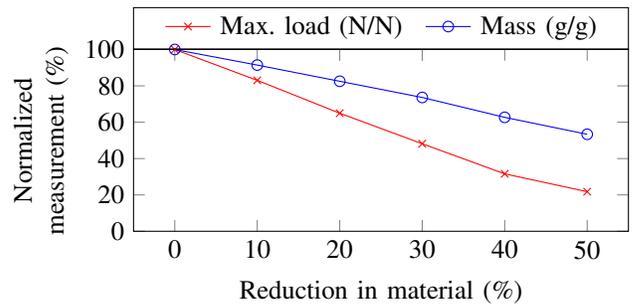
\begin{figure}[h]
    \vspace{-3mm}
    \centering
    \begin{tikzpicture}
    \pgfplotsset{set layers}
        \begin{axis}[name=border,
        	xlabel={Reduction in material (\%)},
        	ylabel style={align=center, text width=3cm},
        	ylabel={Normalized measurement (\%)},
            ymin=0,ymax=100,
        	width=0.45\textwidth,height=4cm,
            ]
        
        \addplot[color=red,mark=x] coordinates {

       	(0, 100)
       	(10, 83.03)
       	(20, 64.95)
       	(30, 48.15)
       	(40, 31.6)
       	(50, 21.78)
        }; \label{pgfplots:global-extrude-delete}
        
        \addplot[color=blue,mark=o] coordinates {
        (0, 100)
       	(10, 91.43)
       	(20, 82.52)
       	(30, 73.62)
       	(40, 62.68)
       	(50, 53.36)
        
        };\label{pgfplots:global-mass-delete}
        
        \end{axis}
        \pgfextractx{\tempdima}{\pgfpointdiff{\pgfpointanchor{border}{west}}{\pgfpointanchor{border}{east}}}
\addtolength{\tempdima}{-.666em}
\node[draw,inner sep=.333em,above] at (border.north)
  {\makebox[\tempdima]{\ref{pgfplots:global-extrude-delete} Max. load (N/N)\hfil\ref{pgfplots:global-mass-delete} Mass (g/g)}};
        
        \end{tikzpicture}
    \caption{\textit{Printer A} - Max. tensile load vs material reduction.}
    \label{fig:pa-material-reduction}
\end{figure}

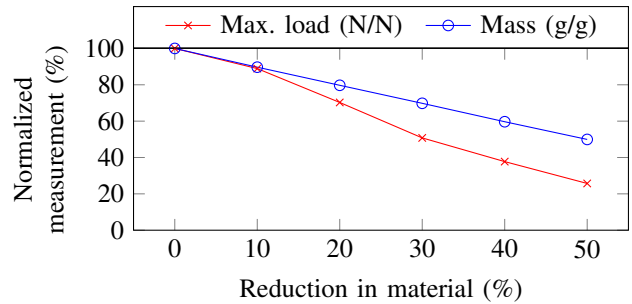
\begin{figure}[h]
    \vspace{-3mm}
    \centering
    \begin{tikzpicture}
    \pgfplotsset{set layers}
        \begin{axis}[name=border,
        	xlabel={Reduction in material (\%)},
        	ylabel style={align=center, text width=3cm},
        	ylabel={Normalized measurement (\%)},
            ymin=0,ymax=100,
        	width=0.45\textwidth,height=4cm,
            ]
        
        \addplot[color=red,mark=x] coordinates {

       	(0, 100)
       	(10, 88.92)
       	(20, 70.28)
       	(30, 50.78)
       	(40, 37.74)
       	(50, 25.72)
        }; \label{pgfplots:global-extrude-delete}
        
        \addplot[color=blue,mark=o] coordinates {
        (0, 100)
       	(10, 89.69)
       	(20, 79.72)
       	(30, 69.81)
       	(40, 59.68)
       	(50, 49.94)
        
        };\label{pgfplots:global-mass-delete}
        
        \end{axis}
        \pgfextractx{\tempdima}{\pgfpointdiff{\pgfpointanchor{border}{west}}{\pgfpointanchor{border}{east}}}
\addtolength{\tempdima}{-.666em}
\node[draw,inner sep=.333em,above] at (border.north)
  {\makebox[\tempdima]{\ref{pgfplots:global-extrude-delete} Max. load (N/N)\hfil\ref{pgfplots:global-mass-delete} Mass (g/g)}};
        
        \end{tikzpicture}
    \caption{\textit{Printer B} - Max. tensile load vs material reduction.}
    \vspace{-3mm}
    \label{fig:pb-material-reduction}
\end{figure}

\subsection{Trojan Attack 2: Material Relocation}
In the \texttt{g-code}, the extrusion values are presented within an \textit{absolute} frame of reference. 
This means that if one extrusion is removed, the next extrusion will deposit extra material to keep the values consistent. Consider three back-to-back commands \texttt{(G1 X1 Y2 E3)}, \texttt{(G1 X2 Y3 E4)}, and \texttt{(G1 X3 Y4 E5)}.  The total extruded material is 3mm after the first command, 4mm after the second (it deposits 1mm), and 5mm after the third.  If the Trojan alters the second command to \texttt{(G0 X2 Y3)}, no material is extruded during its execution, although the head continues along the same route. Crucially, the third command now deposits 2mm and the total material used remains 5mm.

\begin{figure}[b]
    \vspace{-3mm}
    \centering
    \includegraphics[width=0.48\textwidth]{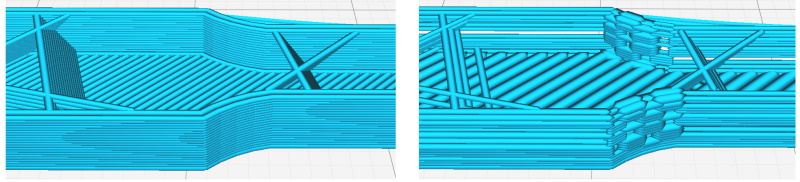}
    \caption{Simulation of (L) original and (R) 1-in-2 relocation.}
    \label{fig:vis-relocation}
\end{figure}

This is the basis of the second attack, which scans for \texttt{G1} linear movement commands with extrusions, and converts a subset (either 1-in-4, 1-in-3, or 1-in-2) into \texttt{G0} linear movement commands with no extrusions.
To further reduce the visibility of the attack, we also add a new activation trigger: we preclude the Trojan from activating until 25\% of the part is printed, and deactivate it after 75\% is printed. For this, we track the \texttt{M73} commands which are used to update the percentage remaining on printer displays. Fig.~\ref{fig:vis-relocation} shows the consequences on a specimen cross section. 

\begin{figure}[t]
    \centering
    \begin{tikzpicture}
\begin{axis}[name=border,
    ybar,
    height=4cm,
	enlargelimits=0.15,
    width=0.48\textwidth,
    legend style={at={(0.5,-0.3)},
      anchor=north,legend columns=-1},
    ylabel={Normalized max. load (\%)},
    ylabel style={align=center, text width=3cm},
    ymin=50,
    x tick label style={font=\scriptsize,rotate=45,anchor=east},
    bar width=3.5mm,
    symbolic x coords={Control,1 in 4, 1 in 3, 1 in 2},
    xtick=data,
    nodes near coords,
    nodes near coords align={vertical},
    every node near coord/.append style={font=\scriptsize,rotate=90,anchor=east}
    ]

\addplot coordinates 
{(Control, 100) (1 in 4,98.85) (1 in 3,94.68) (1 in 2,93.82)};\label{pgfplots:pa}
\addplot coordinates 
{(Control, 100) (1 in 4,97.71) (1 in 3,88.44) (1 in 2,79.77)};\label{pgfplots:pb}

\path (axis cs:Control,0) -- coordinate (m) (axis cs:1 in 4,0);
\draw [dashed] (m) -- (current axis.north -| m);

\end{axis}
\pgfextractx{\tempdima}{\pgfpointdiff{\pgfpointanchor{border}{west}}{\pgfpointanchor{border}{east}}}
\addtolength{\tempdima}{-.666em}
\node[draw,inner sep=.333em,above] at (border.north)
  {\makebox[\tempdima]{\ref{pgfplots:pa} Printer A\hfil\ref{pgfplots:pb} Printer B}};
  
\end{tikzpicture}
    \caption{Material relocation Trojan.}
    \label{fig:pla-relocation}
\end{figure}
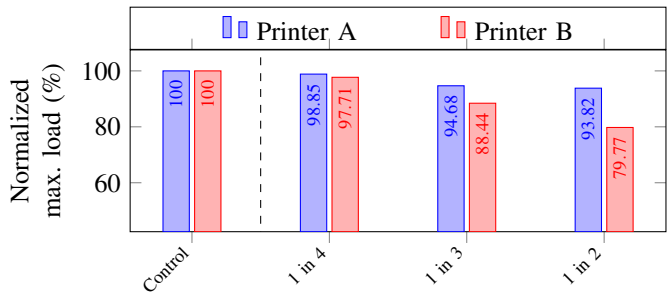

\begin{figure}[t!]
    \vspace{-3mm}
    \centering
    \includegraphics[width=0.46\textwidth]{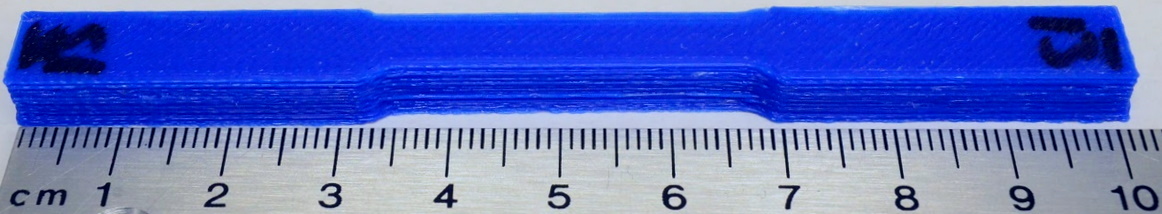}
    \caption{Example 1-in-2 material relocation. No obvious changes are visible w.r.t. specimen quality.}
    \label{fig:material-relocation}
    \vspace{-4mm}
\end{figure}

The results of this attack are depicted in Fig.~\ref{fig:pla-relocation}. All printed objects remain within 1-2\% of the control masses, and despite the edits, the parts are visually similar (example: Fig.~\ref{fig:material-relocation}).
While all edits weakened the parts, the attack on \textit{Printer B} was more effective than the attack on \textit{Printer A}, which is likely due to the different slicing infill strategies (10\% compared to 18\%).
This attack increases the bootloader size to 6986 bytes, a \textbf{1200 bytes} increase, smaller than the previous attack due to the simpler string manipulations even with \texttt{M73} tracking.

\subsection{Discussion}
Different attacks can change the strength profiles of printed parts in different ways. Reducing the printed material is simple and reliable, but noticeable (e.g. via weight). Material relocation is harder to detect, but less consistent, as evident from the differing effectiveness on \textit{Printer A} compared to \textit{Printer B}. 
Given its small size, ($<$1.4KiB) the attack surface for such Trojans within AM CPS is enormous,  since a large number of microcontrollers are present in complex `commercial-grade' systems which have large feature sets.
In addition, the Trojan could be made elusive by altering its trigger. While in this study the attack activated in every print, it would be insidious to activate rarely, with the intent of lowering the average quality of service / institutional reputation.  Parts that are unpredictably faulty can get through certain quality controls, and if targeted at critical products could have catastrophic consequences (for instance consider a Trojan within 3D printers for aerospace components~\cite{nickels_am_2015} or bespoke medical parts~\cite{placone_recent_2018}).

\subsection{Starving \sol}
\label{sec:prevention}
Since this is a bootloader Trojan, it is difficult to detect at the software level -- especially since it has the ability to alter high-level firmware prior to its installation.
In other words, even if the defenses are encoded within user applications, they could be detected and disabled prior to activation.

\begin{figure}[t]
    \centering
    \includegraphics[width=0.4\textwidth]{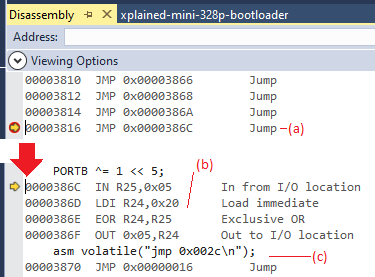}
    \caption{Debugger observing code injection ISR}
    \label{fig:visible_jump}
    \vspace{-4mm}
\end{figure}

However, as \sol\ alters the printer's behaviour, it is not impossible to detect.
Side channels can be monitored with external devices such as cameras and microphones from an arms length -- though this would be intricate as the Trojan effects extrusions and not the head movement.
Careful measurement of the ISR delays could reveal the presence of injected instructions. External programming tools could be utilised -- if monitored with a debugger, the alternate ISR jumps can be noticed.
An example of this is presented in Fig.~\ref{fig:visible_jump}, which shows how the Atmel Studio debugger for an ATmega328P can be used to deduce a bootloader performing a code injection using the methodology in \S\ref{sec:trojan-generic}. Here the Trojan is designed to flip the fifth bit of \texttt{PORTB} whenever the target interrupt occurs. Within the debugger, we can observe that the assembly at (a) is the interrupt target, which then jumps to the injection exploit at (b), flipping a bit in \texttt{PORTB} before the correct ISR is jumped to at (c).

That said, in order to perform this test the embedded system must be designed to allow access to the AVR's debugging (JTAG) port, and toolchains must be utilized that support first-class debugging (e.g. not the Arduino IDE, which offers no live debugging facilities).
This access is not guaranteed, especially given the design constraints of the chosen AVR device and embedded system.
For instance, on the ATmega2560, the JTAG pins are multiplexed with four of the ADC inputs. This means that when the ADCs are in use then the JTAG port is rendered unavailable. This is the case in \textit{Printer A}.
An alternative that does not require the debugging port is to export the bootloader itself from the flash memory of the AVR. As this is performed using the programming hardware of the AVR chip itself, there is no way for the bootloader to edit itself before download. Then, the bootloader may be disassembled / decompiled and audited for malicious behaviour, though this will require specialist tools and knowledge.

As low-end microcontrollers do not support features such as secure boot, further prevention within the hardware can come if they are exchanged for something with more features. Policies surrounding firmware installation/inspection (including bootloaders) could be introduced to mitigate the attack vectors in \S\ref{sec:attack-surface}.
Overall  this work motivates the inclusion of a trusted execution environment~\cite{sabt_trusted_2015} within AM printers.







\section{Conclusions}
\label{sec:conclusions}

The cybersecurity of \ac{AM} CPS is important. We examine a case study in detail, presenting the design space exploration of a bootloader Trojan, including (a) methodology for inclusion, (b) evasion of detection, (c) trigger customization, and (d) malicious payload.
Even within tight constraints (both attacks less than 1.7KB in size!), we were able to craft attacks which lowered the strength of printed parts by up to 50\%.
Though we frame the issue around desktop AM devices, we stress that the issues we highlight in this paper are not restricted to these models.
Indeed, the more complex an AM CPS is, the greater the attack surface for embedded Trojans, and `commercial grade'/`industrial scale' 3D printers have complex internal networks of microcontrollers and embedded systems.
This work serves as a reminder that these components can hide malicious surprises, especially when they support complex and powerful configuration options that can be misused.
It takes only one component to be infected by a malafide insider or malicious third party with access to cause insidious and catastrophic consequences.
We believe that procedures for bootloader and firmware verification should be introduced across the AM CPS space, alongside potential automatic monitoring (e.g. via side channels) which could be developed to detect and flag anomalous behaviour.


\section*{Acknowledgments}

This work was supported in part by National Science Foundation SaTC-EDU grant DGE-1931724. We also thank Gary Mac for his help with the CAD modelling.


\ifCLASSOPTIONcaptionsoff
  \newpage
\fi

\bibliographystyle{IEEEtran}
\bibliography{Hack-Anet-A8-v2}

\end{document}